\documentstyle[epsf]{article}

\newcommand{\be}{\begin{equation}}
\newcommand{\ee}{\end{equation}}

\begin{document}

\begin{flushright}
Liverpool Preprint: LTH 620\\
 \end{flushright}

  
\vspace{5mm}
\begin{center}
{\LARGE \bf Hadronic decay of a scalar B  meson from the lattice}\\[10mm] 
{\large\it UKQCD Collaboration}\\[3mm]
 
 {\bf C. McNeile,   C.~Michael and G. Thompson \\
Theoretical Physics Division, Dept. of Mathematical Sciences, 
          University of Liverpool, Liverpool L69 3BX, UK 
 }\\[2mm]

\end{center}

\begin{abstract}

 We explore the transitions B$(0^+)$ to B $\pi$ and  B$_s(0^+)$ to B K 
from lattice QCD with $N_f=2$ flavours of sea quark, using the static
approximation for the heavy quark. We evaluate the effective coupling
constants, predicting  a B$(0^+)$ to B $\pi$ width  of around 160 MeV.
Our result for the coupling strength  adds to the evidence that the 
B$_s(0^+)$ meson is not predominantly a molecular state (BK).

\end{abstract}

\section{Introduction}

 The interest in the excited states of heavy-light mesons has  been
enhanced by  the striking discovery that the $c\bar{s}$ states with
$J^P=0^+$ and $1^+$  have very narrow
widths~\cite{Aubert:2003fg,Besson:2003cp,Krokovny:2003zq,Abe:2003zm}.
This raises the question of whether the  corresponding   $b\bar{s}$
states will also be narrow. The main  reason for the narrow width of
D$_s$ mesons is that the  transition to DK  is not energetically allowed
(for the 2317 MeV state) or the state is  close to threshold (for the
2457 MeV state). Thus the only  allowed hadronic decay proceeds via
isospin-violation (since $m_d \not = m_u$)  to D$_s$$\pi$ and will have
a very small width. 
 Likewise, if the equivalent $b \bar{s}$ states are close to or  below
the $BK$ threshold, then they will be very  narrow. 

 Lattice studies have addressed the energies of these  P-wave $b
\bar{s}$ states~\cite{M+P,bali,bs} and concluded that they indeed lie
close to  or below threshold and hence have very small decay widths.
Although the  lattice studies use $b \bar{s}$ creation operators for
these states, it is also possible that a molecular description  (as a BK
bound state) is more appropriate, as has been suggested for the $c
\bar{s}$ case~\cite{bcl}. To clarify this situation further,  it would
be very useful to evaluate the hadronic transition strength  from the
scalar B state to a B meson plus a light pseudoscalar meson. 

 Here we evaluate these hadronic transition amplitudes using lattice
methods. This has relevance to the decay of a scalar B or B$_s$ meson
to B plus a pseudoscalar meson.

 \section{Spectrum}

   In the heavy quark limit, the $\bar{Q} q$ meson, which we refer to as
a `$B$' meson, will be the `hydrogen  atom' of QCD. Since the meson is
made from non-identical quarks, charge conjugation  is not a good
quantum number.  States can be  labelled by  L$_{\pm}$, where the
coupling of the light quark spin to the orbital angular momentum  gives 
$j_q=L\pm {1 \over 2}$. In the heavy quark  limit these states will be
doubly degenerate since the heavy quark spin interaction can be 
neglected, so  the S state  will have $J^{PC}=0^-,\ 1^-$ while the 
P$_{-}$ state will have  $J^{P}=0^+,\ 1^+$. Note that in the static
case, the  self energy of the static heavy quark is unphysical, so that
only mass differences are physical. Our notation for these static-light 
mesons is B($nL_{\pm}$) where $n=1$ (often omitted) is the ground state, 
and $n=2$ the first excited states, etc.
 Here we will be studying the transition from the P$_{-}$ state to the S
state emitting a pion  in a relative S-wave. This can be applied to the
decays $B(0^+)  \to B(0^-) \pi$ and to  $B(1^+)  \to B(1^-) \pi$.

 We shall be using the $N_f=2$  lattice configurations~\cite{ukqcd}
with $\beta=5.2$ and volume $16^3 \times 32$ with SW-clover improvement 
coefficient 2.0171.  We only use the unitary points, namely 
those with valence light quarks of the same mass as the sea quarks.  The
details of the spectrum from ref.~\cite{bs}  are collected in
Table~\ref{tab_df}.



 The method we shall use to obtain 3-point correlations (next section)
using  timeslice random sources can be used for 2-point  correlations
and  compared with the maximal variance reduction (MVR) 
method~\cite{M+P} used for the 2-point correlators in extracting the 
spectrum~\cite{bs}. For our lighter quark mass, we find the local-local
B(S) correlator  is more precisely determined for $t > 4$ by 40 gauge
configurations of MVR  than 100 gauge configurations of time-slice
evaluation, although the latter had  a somewhat smaller computational
overhead. Since larger $t$ is  important for separating ground states
and excited states, MVR is the  method of choice for the 2-point
correlation. Because it does not  generalise efficiently to the 3-point
correlation, we use the timeslice method  there.


\begin{table}

 \begin{center}
 \begin{tabular}{|l|ll|}
 \hline      
 $\kappa$            & 0.1355    & 0.1350        \\
  MVR gauges     & 40            & 20            \\
  t slice  gauges  & 100            & 20            \\
 $r_0/a$             & 5.041(40)     & 4.754(40)       \\
 $r_0 m(0^{-+})$     & 1.48(3)   & 1.93(3)      \\
 \hline
 $r_0 m(1S)$              & 3.73(8)     &  3.68(7)       \\
 $r_0 m(2S)$              & 5.60(14)    &  5.61(8)      \\
 $r_0 m(1P_{-})$          & 4.75(6)     &  4.71(8)     \\
 $r_0 m(2P_{-})$          & 7.38(9)     &  7.1(2)     \\
 \hline   
\end{tabular}       
 \end{center}

 \caption{ Lattice parameters and results from ref.~\cite{bs} for the
energies of $Q\bar{q}$ states in units of $r_0$ for dynamical fermions
with $N_f=2$. The values of $r_0/a$ and the $q \bar{q}$ pseudoscalar
meson mass are from ref.~\cite{ukqcd}.  Here we set the scale using 
$r_0$ of 0.525(25) fm. The heavy-light meson lattice energies contain
the static source self-energy so that only differences are physical.}
 \label{tab_df}
\end{table}

 \section{Decay transitions}

 Following the methods~\cite{hdecay,rhodecay} used to study the hadronic
transitions such as hybrid decay and $\rho$ to $ \pi \pi$, we can
determine the transition amplitude  provided that there is approximate
equality of energies  between the initial state B(1P$_{-}$) and the final
two-body state B(1S) $\pi$.  Here we are taking the $b$ quark as
static and using two flavours  of light quark. Staying within the fully
unitary sector of the theory,  we can study transitions with the same
valence quarks in the  B mesons and pion as in the sea.

\begin{figure}[th]
 \centerline{\epsfxsize=8cm\epsfbox{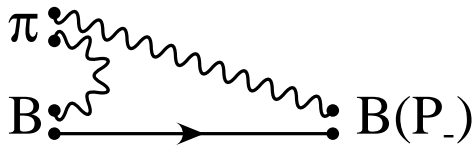}}  
 \caption{ Quark diagram evaluated for the transition from  $B(P_{-})$ 
to B $\pi$. Here $B(P_{-})$ is the static-light meson that comprises
degenerate scalar and axial $B$ mesons.
 }
 \label{BsBpi.fig} 
 \end{figure}

 The lightest two-body  B(1S) $\pi$ state on a lattice will be when the
pion has relative momentum zero. The energy differences are then given
by $a \Delta E= 0.09,\ 0.19$ for light quarks of mass corresponding
approximately to  $2m_s/3$ and $m_s$ for $\kappa=0.1355$ and 0.1350
respectively~\cite{bs}. Especially for the lighter  quark case, this is
an energy difference small enough to apply the
method~\cite{hdecay,rhodecay}, namely
 $\Delta E t \ll 5$, up to  large $t$-values.

 We need to evaluate correlations between B(1P$_{-}$) at $t=0$ and B(1S)
$\pi$ at time $t$. This involves quark propagators between three
space-time points. This is illustrated in fig.~\ref{BsBpi.fig}. The
heavy quark propagator, however, is trivial to evaluate: as a product 
of gauge links with a $(1 \pm \gamma_4)/2$ projector for spin. We create
 B(S) as $Q \gamma_5 \bar{q}$ and  B(P$_{-}$) as $Q \bar{q}$,  and, in
both cases, also two different fuzzed versions of these~\cite{M+P,bs}.
In this  exploratory study, we only consider a pion with zero momentum
(so we  sum over relative spatial position) with a local creation
operator $q \gamma_5 \bar{q}$.

 To gain sufficient statistics for the three point correlations, we wish
to  evaluate the correlation using  every space and time point  on the
lattice as a source.   To achieve this,   we follow the stochastic
technique used previously~\cite{fm,bs}. We use  a stochastic source
$\xi$(complex gaussian random number in every colour,  dirac, space
component) at a given time slice $t$. We then evaluate the  propagator
$\phi$ from this source using $M \phi = \xi$ where  $M$ is the
Wilson-Dirac matrix for the light quark. The required correlation can
then be  obtained from this propagator, schematically as, 
 \be
 C_3(t'-t)=  \phi^*(t',y)_{ai} \phi(t,y)_{bj} (1+\gamma_4)_{ji} 
 \left( \Pi_{t''=t,t'-1} U(y,t'') \right)_{ab}
 \ee
 where all repeated indices (and $y$) are summed. The product of
stochastic sources implicit in the product of $\phi$'s  then  gives an
expectation value which is the required sum  over pion sources
throughout the time-slice at $t$, whereas the  noise terms average to
zero. By using more independent stochastic samples, the average over 
them (from eq.~1) will have reduced noise. One can also improve the 
signal to noise ratio by combining results from different time values
$t$ for the  stochastic source.

 In practice we used one stochastic sample per time slice, but all time
slices in turn, as in ref.~\cite{rhodecay}. This implies  32 inversions
per gauge to evaluate the required 3 point correlation from  all sources
to all sinks. This is computationally very efficient and provides
sufficient  precision, as shown below. We use 100 gauges for the lighter
 quark mass and 20 for the heavier.

The motivation for using a source restricted to  one time-slice is
to ensure that the noise contributions decrease  as the signal decreases
with increasing $|t'-t|$. Note that a two point correlator, for  example
for the pion with zero momentum and local creation and destruction
operator,  can be obtained likewise from
 \be
 C_{\pi}(t'-t)=  \phi^*(t',y)_{ai} \phi(t',y)_{ai} 
 \ee

 In this work, we extract the ground state pion contribution to $C_{\pi}$
 from a fit to pion correlations obtained from conventional
analyses~\cite{ukqcd} with non-stochastic sources, so  we  do not use
the stochastic result for the 2-point pion correlators, other than as a
check.

 Likewise, for B(P$_{-}$) with local creation and destruction operators,
 the two-point correlator can be obtained from
 \be
 C_{B(P_{-})}(t'-t)=  \phi^*(t',y)_{ai} \xi(t,y)_{bj} (1+\gamma_4)_{ji} 
 \left( \Pi_{t''=t,t'-1} U(y,t'') \right)_{ab}
 \ee  
 As we discussed above, this latter expression for $C_B$ is more noisy
than the  MVR method at larger $|t'-t|$, so we again only use it as a
cross check.

\begin{figure}[th]
 \vspace{-2cm}
 \centerline{\epsfxsize=11cm\epsfbox{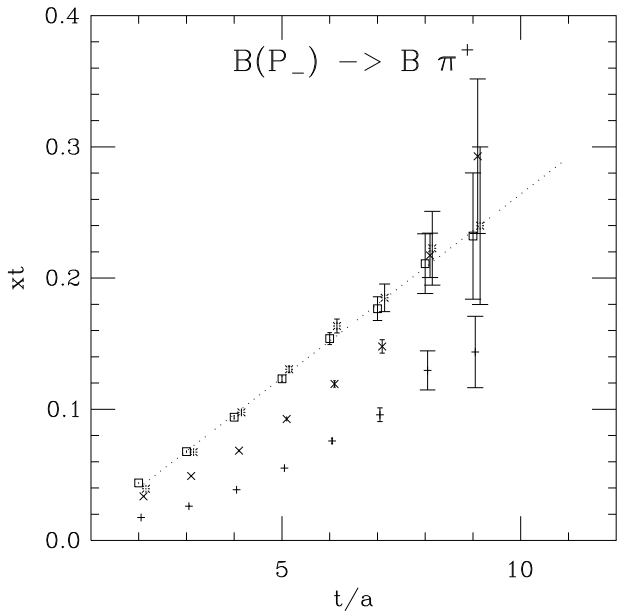}}  
 \caption{ Normalised three  particle correlator  versus $t/a$ 
for $\kappa$=0.1355. The points marked ($+,\ \times,\ *$) are for 
local, lightly fuzzed and heavily fuzzed operators 
respectively. The combination which optimises the ground 
state is shown by squares, and a linear fit to it is shown.
 }   
 \label{xt.fig} 
 \end{figure}

 The normalised transition amplitude $x$ on a lattice can then be 
obtained from the ratio
 \be
 { C_3(t) \over \sqrt{ C_{B(P_{-})}(t) C_{\pi}(t)  C_{B(S)}(t) }  }
 = xt + {\rm const}
 \ee
  provided that the transition rate is not too large, namely $xt \ll 1$.
 This ratio for the decay to $\pi^{+}$ is plotted for  our lighter quark
mass in fig.~\ref{xt.fig}. As well as illustrating  the result for each
of our three operators to create a heavy-light meson, we can choose to
improve the  ground state projection of the B(S) and B(P$_{-}$) by using
an appropriate  linear combination  of local and fuzzed operators. The 
ratio for this improved projection is also illustrated.

\begin{figure}[th]
 \vspace{-2cm}
 \centerline{\epsfxsize=11cm\epsfbox{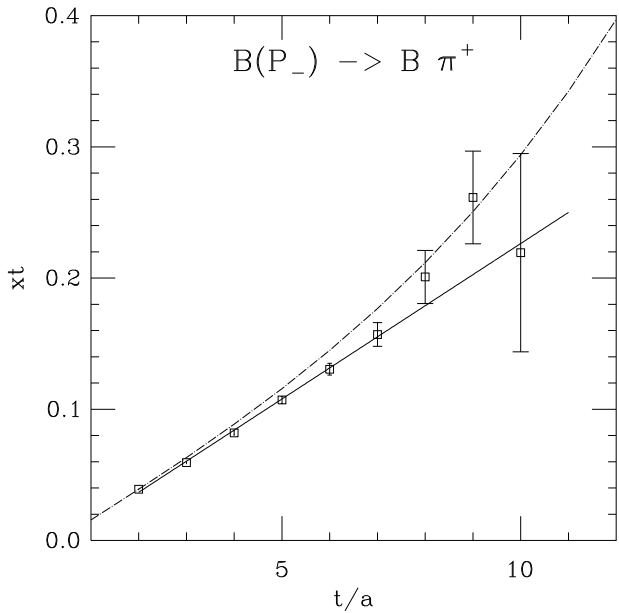}} 
 \caption{ Normalised three  particle correlator  versus $t/a$  for
$\kappa$=0.1350.  The combination which optimises the ground  state is
shown together with a linear fit. We also show the  result of a two
state model with the correct energy difference ($a\Delta E=0.19$)
between  the two body and scalar B states.
 }   
 \label{xt0.fig} 
 \end{figure}

This result shows a linear behaviour, as expected if excited state
contributions  are not significant. We can then read off the hadronic
transition amplitude  $ax$ from the slope - obtaining $ax=0.028(3)$.

 This is the transition with lattice normalisation and for one quark
diagram.  For the transition B$(P_{-}) \to$ B(S) $\pi$, there will be
two quark diagrams  contributing (since either a $u$ or $d$ quark pair
can be produced, yielding $\pi^{+}$ or $\pi^{0}$)  and the overall  rate
will be 3/2 that evaluated from the amplitude $x$ above. To derive the
appropriate normalisation~\cite{hdecay}, consider the decay width, even
though  the decay is not energetically allowed with our parameters. Then
$\Gamma=2 \pi x^2 \rho$ where $\rho$ is the two body  phase space which
evaluates to $\rho=L^3 k E_{\pi} /(2 \pi^2)$. We have an isotropic decay
(S-wave) and it is reasonable to assume that  $x$ is independent of the
decay momentum $k$.

To increase predictive power,  we evaluate the  coupling constant in an
effective lagranian for the three point vertex  describing the decay.
This coupling constant has the dimensions of mass, so  we set the scale
using the mass of the decaying meson. Then the coupling constant 
squared is proportional to $\Gamma/k$, and we use that definition as an 
effective coupling strength. We then obtain an effective coupling
strength  of 
 \be
     \Gamma/k=3 (L/a)^3 (ax)^2 a E_{\pi}    /(2 \pi) = 0.46(9)
 \ee
 where for the pion at zero monentum, we may use its lattice
mass~\cite{ukqcd,k1358} for $E_{\pi}$.

 In order to explore the dependence on the light quark mass, we use
another quark mass, although we are limited by the need to keep the 
transition approximately on mass-shell, so with decay products of
similar energy to  the initial scalar meson.  We used  $\kappa=0.1350$
where the  quark mass is approximately strange.  We find a similar plot
(fig.~\ref{xt0.fig}) of $xt$ versus $t/a$ with a slope  of 0.0237. Since
in this case we have a somewhat bigger  mismatch (namely $a \Delta E =
0.19$) between the energies for the two body state and of the  scalar
meson, we can correct for this by using a two state
model~\cite{rhodecay}. This shows that we would expect  some small
curvature, even in the ideal case when there  is no contribution from
excited states. Our lattice result is  quite consistent with this
curvature. Because of the additional analysis  needed to cope with the
larger energy gap, the systematic errors  are relatively larger in this
case. We estimate $ax=0.024(4)$. Then the effective coupling strength
is  $\Gamma/k=0.46(9)$, exactly the same value as obtained at the lighter 
quark mass.

Since a  quark-antiquark pair is created in the decay, it might be 
expected that the amplitude to produce heavier quarks was smaller.
However, a major component of the transition amplitude may come
from  considerations of the overlap of the initial and final states, and
this  does not depend on the light quark mass in any very simple way. 
Indeed in  our study~\cite{rhodecay} of $\rho$ decay to two pions, we
saw some evidence that the  decay amplitude was largely independent of 
the light quark mass.  This is what we find here for scalar decays also.

 The method we have used to evaluate the hadronic transition is only 
approximate, and assumes that the transition amplitude $x$ is relatively
small. Since we find that $xa \approx 0.03$, this is indeed justified. 
 In general, however, one can proceed in a rigorous way. This involves 
determining the energy of the two body system (B$(1S)+ \pi$) as
accurately as  possible with a full QCD lattice simulation, and then
obtaining the  dependence of this on the lattice spatial volume. This
then  gives information on the scattering phase shift in the two body
channel~\cite{Luscher:1985dn,Luscher:1986pf,Luscher:1990ux,Luscher:1991cf}. 
It is, of course, consistent  to treat the static
quark as quenched, but all the light quarks need to be  treated
dynamically. In our approach with $N_f=2$ flavours of degenerate  sea
quark, this would allow a study of the transition from B$(1P_{-})$  to
B$(1S)+ \pi$. To have the most accurate determination of the two-body 
energy, one should use a variational approach with both two body and 
one body operators. This will involve the three body correlation we have
measured above, but also the two to two and box quark diagrams. Our
preliminary study  indicates, as was found for the case of $\rho$
decay~\cite{rhodecay},  that these diagrams are too noisy to yield
sufficiently accurate results, even measuring from all space-time
sources for 100 gauge configurations.

 \section{Discussion}

 As discussed above, we are able to measure the transition amplitude 
from  the $0^+$ $b\bar{q}$ meson to B $\pi$, provided that the  light
quark masses are such that the initial and final states have  very
similar energies. For the case we have explored, with $N_f=2$ flavours
of degenerate light quark, this implies that we must extrapolate in the
light quark mass  to make contact with experiment. This we do by 
assuming that the  coupling constant for the transition,  as described
by an  effective lagrangian, is independent of the light meson mass.
This leads to the assumption that  the effective coupling strength
$\Gamma/k$ introduced above  will be independent of the light quark
mass. We do indeed see some evidence from our lattice results that this
is the case.
 Thus we shall use our lattice results  for the reduced width, evaluated
where  no decays are allowed,  to compare with experiment and to make
predictions. Since we work at a fixed lattice spacing, we are unable to 
estimate the systematic error arising  from not taking the continuum
limit.

There is a state known experimentally~\cite{pdg} which  is a candidate
for the $0^+$ $b\bar{q}$ meson, namely the  B$^{**}$  with mass 5698(8)
MeV and width 128(18) MeV. This  corresponds to an  effective coupling
strength of $\Gamma/k=0.34(5)$. However, the experimental state may be
a superposition of several states, so mass values and widths  for the
$0^+$ state are not really known experimentally. 


 From lattice studies with static quarks, the excitation energy  of the
scalar B$^{**}$ state is estimated~\cite{bs} to be $368\pm31$ MeV, 
where this energy difference was evaluated for strange light quarks, but
 was expected to be similar for non-strange light quarks. Using this 
central value of 368 MeV for the energy release, the width of the 
scalar  B$^{**}$ state, with decay to B$\pi$, would be 162(30)  MeV.  
 Our result is significantly lower than that  
obtained~\cite{Bardeen:2003kt,Nowak:2003ra} 
using a chiral symmetry between the $0^{\pm}$ B mesons, namely  a width
of around 500 MeV using $G_A \approx 1$.


 It may also be relevant to compare with experimental data on decays  of
heavy-light mesons with charm quarks, since there is a wider range of
data
available~\cite{Aubert:2003fg,Besson:2003cp,Krokovny:2003zq,Abe:2003zm,pdg}.
 From the  observed~\cite{Abe:2003zm} mass of  2308$\pm17\pm15\pm28$ MeV
for D(0$^+$) and width of 276$\pm21\pm18\pm60$ MeV for  decay to
D(0$^-$) + $\pi$, one gets $\Gamma/k = 0.73_{-24}^{+28}$.   This is a
somewhat larger  effective coupling strength than the value of  0.46(9)
that we obtained above (but consistent within errors), although our
evaluation is  for static quarks whereas charm quarks are known to be
sufficiently  light that this can be a poor approximation for them. 


It is also possible to extract an effective coupling strength for the 
decay of K(1412) to K $\pi$, obtaining~\cite{pdg}  $\Gamma/k =0.48(5)$.
Thus the experimental data are consistent with an effective coupling
strength  of about 0.5 for decays of scalar heavy-light mesons with
heavy quarks  that are $b,\ c$ and $s$. This is very consistent with our 
{\it ab initio}  evaluation which gives around 0.5 also.

 For the $b \bar{s}$ excited mesons, in the limit of degenerate $u$  and
$d$ quarks, there will be no decay to pions and the main hadronic  decay
will be to BK with the emission of a light quark-antiquark pair. In this
case our evaluation is partially  quenched, in the sense that the
strange quark in the B$_s$ meson  and K meson is not present in the sea.
For the decay of a scalar B$_s$ meson,  the energy release may be small
or  the state may even be stable~\cite{bs}. Even if the state is  stable
under strong interactions, we can still evaluate the hadronic transition
strength  as an effective coupling. Consider the transition B$_s(P_{-})
\to$ B(S) K,  there are again two quark diagrams, now with equal weight.
Our result is then that $\Gamma/k = 0.61(12)$. 
 If this scalar meson does lie above threshold,  we predict a width given
by that expression.

Consider now  whether the  B$_s(P_{-})$ meson is a quark-antiquark state
or a BK meson. Since we have found a non-zero transition amplitude (our
$x$)  on a lattice it follows that the meson and the two-body state mix.
Indeed  when the meson mass is degenerate with the two-body energy,
there will be an  avoided level-crossing, with full mixing. What is more
significant, however, is the  situation in a large volume, when the two
body energy spectrum becomes  continuous.

  The situation in lattice studies is then  more like in experiment -
one has to deduce the composition of a hadron from  its observed
properties. There are lots of extra clues  available in lattice studies,
however: (i) the mass of the state can be explored as  the quark mass
varies, (ii) the wave-function and charge form factor of the state can
be measured,  (iii) the coupling strength of transitions can be
evaluated. For the  B$(P_{-})$ meson, lattice studies with $N_f=2$ show
a spectrum~\cite{bs} with a mass which is more-or-less independent of
the  two body (B$\pi$) threshold  which would not  be expected for a
molecular state. The coupling strength, as discussed above,  is similar
to that  for the scalar decay  of K(1412) to K $\pi$, where a molecular
structure is not expected. The  charge form factor has only been
measured~\cite{wf} for $N_f=2$ for the ground state B(S) although
quenched results~\cite{M+P} show a Bethe Salpeter wavefunction for 
B$(P_{-})$ similar to quark model expectations. A more definitive
lattice  conclusion must await studies with lighter quarks, but all the
evidence at present  points to the heavy-light scalar meson as not being
predominantly a molecular state.


\end{document}